\documentclass[aps,prl,twocolumn,groupedaddress]{revtex4-1}
\usepackage{graphicx}
\usepackage{tikz}
\usetikzlibrary{arrows,shapes}
\usepackage{color}
\usepackage{amsmath}
\usepackage{amssymb} 
\usepackage{array}

\newcommand{\kB}{k_{B}}
\newcommand{\betaconfeqh}{\hat{\beta}_{\rm eqh}}
\newcommand{\betaconfeqa}{\hat{\beta}_{\rm eqa}}
\newcommand{\betatexteqh}{\beta_{\rm eqh}}
\newcommand{\betatexteqa}{\beta_{\rm eqa}}

\newcommand{\tauEff}{\tau_{\rm eff}}

\newcommand{\Stext}{    {S}}
\newcommand{\Sconf}{\hat{S}}
\newcommand{\betatext}{    {\beta}}
\newcommand{\betaconf}{\hat{\beta}}

\newcommand{\tautext}{    {\tau}}
\newcommand{\tauconf}{\hat{\tau}}

\begin{document}

\title{
  Canonical free-energy barrier of particle and polymer cluster formation
}
\author{Johannes Zierenberg}
\email[]{johannes.zierenberg@itp.uni-leipzig.de}
\affiliation{Institut f\"ur Theoretische Physik, 
             Universit\"at Leipzig, 
             Postfach 100\,920, 
             D-04009 Leipzig, 
             Germany
           }
\author{Philipp Schierz}
\email[]{philipp.schierz@itp.uni-leipzig.de}
\affiliation{Institut f\"ur Theoretische Physik, 
             Universit\"at Leipzig, 
             Postfach 100\,920, 
             D-04009 Leipzig, 
             Germany
           }
\author{Wolfhard Janke}
\email[]{wolfhard.janke@itp.uni-leipzig.de}
\affiliation{Institut f\"ur Theoretische Physik, 
             Universit\"at Leipzig, 
             Postfach 100\,920, 
             D-04009 Leipzig, 
             Germany
           }
%\affiliation{}

\date{\today}

\begin{abstract}
  %A common approach to study nucleation rates is the estimation of free-energy
  %barriers. This usually requires knowledge about the shape of the forming
  %droplet, a task that becomes notoriously difficult in macromolecular setups
  %starting with a proper definition of the cluster boundary or a proper
  %ensemble choice. Here, we demonstrate that a shape-free determination of
  %temperature-driven cluster formation is directly accessible in the canonical
  %ensemble for particle as well as polymer systems. Combined with rigorous
  %results on equilibrium droplet formation, this allows for a well-defined
  %finite-size scaling analysis of the effective interfacial free energy at
  %fixed density. We first verify the theoretical predictions for the formation
  %of a liquid droplet in a supersaturated particle gas by generalized-ensemble
  %Monte Carlo simulations of a Lennard-Jones system.
  %% addon
  %Going one step further, we then generalize this approach to cluster formation
  %in a dilute polymer solution. Our results suggest an analogy with particle
  %condensation, when the macromolecules are interpreted as extended particles. 
  %% kinetic part
  %Because a standard approach in Monte Carlo simulations is to work in the
  %conformational ensemble, we show that excluding the kinetic energy from the
  %partition function leads to finite-size differences in the free energy but
  %retains intensive parameters in the thermodynamic limit. 
  %
  A common approach to study nucleation rates is the estimation of free-energy
  barriers. This usually requires knowledge about the shape of the forming
  droplet, a task that becomes notoriously difficult in macromolecular setups
  starting with a proper definition of the cluster boundary. Here, we
  demonstrate a shape-free determination of the free energy for
  temperature-driven cluster formation in particle as well as polymer systems.
  Combined with rigorous results on equilibrium droplet formation, this allows
  for a well-defined finite-size scaling analysis of the effective interfacial
  free energy at fixed density. We first verify the theoretical predictions for
  the formation of a liquid droplet in a supersaturated particle gas by
  generalized-ensemble Monte Carlo simulations of a Lennard-Jones system. Going
  one step further, we then generalize this approach to cluster formation in a
  dilute polymer solution. Our results suggest an analogy with particle
  condensation, when the macromolecules are interpreted as extended particles.
\end{abstract}

%\pacs{
%\keywords{}

\maketitle

% potential energy is adequate reaction coordinate for nucleation-like ...
\mbox{}
\clearpage

%%%%%%%%%%%%%%%%%%%%%%%%%%%%%%%%%%%%%%%%%%%%%%%%%%%%%%%%%%%%%%%%%%%%%%%%%%%%%%%
%%%%%%%%%%%%%%%%%%%%%%%%%%%%%%%%%%%%%%%%%%%%%%%%%%%%%%%%%%%%%%%%%%%%%%%%%%%%%%%
%%%%%%%%%%%%%%%%%%%%%%%%%%%%%%%%%%%%%%%%%%%%%%%%%%%%%%%%%%%%%%%%%%%%%%%%%%%%%%%
%%%%%%%% Introduction
The formation of equilibrium droplets from a supersaturated gas is a
long-standing subject of interest, being an essential phase transition in
nature~\cite{feder1966, oxtoby1992, kashchiev2000}. More importantly, the
underlying mechanism is relevant for a multitude of nucleation-like processes
from statistical mechanics to material science. These include crystallization in
colloidal suspensions~\cite{sear2007, auer2001}, cluster formation in protein
solutions~\cite{stradner2004, sear2007}, as well as domain formation in
so-called phase-change materials~\cite{kalb2005, wuttig2009, lee2011} and glassy
solids~\cite{lee2009}. It is even connected to field theory~\cite{langer1967}
and nuclear reactions~\cite{goodmann1984}.
The formal framework of free-energy calculations is straightforward, e.g., in
terms of reaction coordinates in phase space, but the application in computer
simulations is diverse with an ongoing demand for further methodological
developments~\cite{hansen2014}. A seminal application was the parameter-free
estimate of crystal nucleation rates from equilibrium free-energy
barriers~\cite{auer2001}. It seems natural that the estimation of nucleation
barriers becomes increasingly difficult when considering more complex systems
like polymer or protein solutions~\cite{sear2007}. 

% relation to nucleation rates
The rate of nucleation $R$ is related by classical nucleation
theory~\cite{feder1966, kashchiev2000} to the free-energy cost $\beta\Delta F$
of a nucleus on top of the nucleation barrier:
\begin{equation}
  R = \kappa e^{-\beta \Delta F},
  \label{eqRate}
\end{equation}
with the inverse temperature $\beta=1/\kB T$ and the Boltzmann constant $\kB$.
The kinetic prefactor $\kappa$ includes the kinetic details of the nucleation
process, such as diffusion and nucleus-attachment rates. The free-energy barrier
may be related to the suppression in the equilibrium probability distribution.
Physically relevant barriers for liquid-vapor condensation are supposed to be in
the range of $20\kB T$ to $100\kB T$ (see, e.g., Ref.~\cite{schrader2009}). 
The typical setup for the study of free-energy barriers is at fixed temperature
by variation of the density or degree of supersaturation, or directly in the
grand-canonical ensemble. The barrier is then associated to the suppression in
the droplet-size probability distribution~\cite{auer2001} and is clearly
temperature dependent~\cite{hale2010, tanaka2014}. This usually requires
estimating the droplet size and interface, a task that introduces systematic
uncertainties and strongly depends on the droplet definition. Instead, the
free-energy barrier may be directly related to the volume of the critical
nucleus and the pressure difference~\cite{statt2015}, exploiting a thorough
understanding of the underlying phenomenon in a clever way. In this context,
the problem can be reduced to conformational phase space, knowing that
canonical expectation values typically do not depend on the kinetic energy.

In the following, we address the question of how to easily obtain dependable
results on nucleation barriers without invoking elaborate thermodynamic
reasoning or estimating nucleus shapes. This opens the door to more complex
systems with nucleation-like mechanisms such as self-assembly and aggregation,
where the nucleus shapes are a priori unknown. Importantly, we consider a setup
at fixed density with varying temperature -- an intuitive approach from a
condensed matter perspective. We focus on aggregation of polymers in a dilute
setup~\cite{zierenberg2016polymers, zierenberg2015epl, zierenberg2014jcp}
guided by the canonical case of droplet formation in a particle gas.
For the canonical case, we analyze a two-dimensional free-energy landscape and
identify the energy as a suitable reaction coordinate.  This allows us to
formulate the problem in the microcanonical ensemble of either fixed total
energy $E$ or fixed potential energy $E_p$. The first is the usual textbook
definition, while the latter has been frequently applied in recent computer
simulation studies.
%~\cite{janke1998, gross2001, pleimling2001, okamoto2004, junghans2006, chen2008, kastner2009,
% taylor2009, schnabel2011, nogowa2011, zierenberg2015epl, vogel2015, koci2015,
%marenz2016}.
This enables us to directly discuss the effect of kinetic energy when changing
between the two formulations $E\leftrightarrow E_p$. If kinetic energy matters,
only the first one allows a direct physical interpretation. 

%%%%%%%%%%%%%%%%%%%%%%%%%%%%%%%%%%%%%%%%%%%%%%%%%%%%%%%%%%%%%%%%%%%%%%%%%%%%%%%%
%%%%%%%%%%%%%%%%%%%%%%%%%%%%%%%%%%%%%%%%%%%%%%%%%%%%%%%%%%%%%%%%%%%%%%%%%%%%%%%%
%%%%%%%%%%%%%%%%%%%%%%%%%%%%%%%%%%%%%%%%%%%%%%%%%%%%%%%%%%%%%%%%%%%%%%%%%%%%%%%%
%%%%%%%%% Theory: microcanonical ensembles
\section{Results}

%%%%%%%%%%%%%%%%%%%%%%%%%%%%%%%%%%%%%%%%%%%%%%%%%%%%%%%%%%%%%%%%%%%%%%%%%%%%%%%%
%%%%%%%%%%%%%%%%%%%%%%%%%%%%%%%%%%%%%%%%%%%%%%%%%%%%%%%%%%%%%%%%%%%%%%%%%%%%%%%%
%%%%%%%%%%%%%%%%%%%%%%%%%%%%%%%%%%%%%%%%%%%%%%%%%%%%%%%%%%%%%%%%%%%%%%%%%%%%%%%%
%%%%%%%%% Theory: free-energy barrier and scaling
\subsection{Droplet formation free-energy barrier}
\begin{figure}[t!]
  \begin{flushleft}
    $\boldsymbol{E_p}$:
  \end{flushleft}
  \vspace{-2.0em}
  \includegraphics{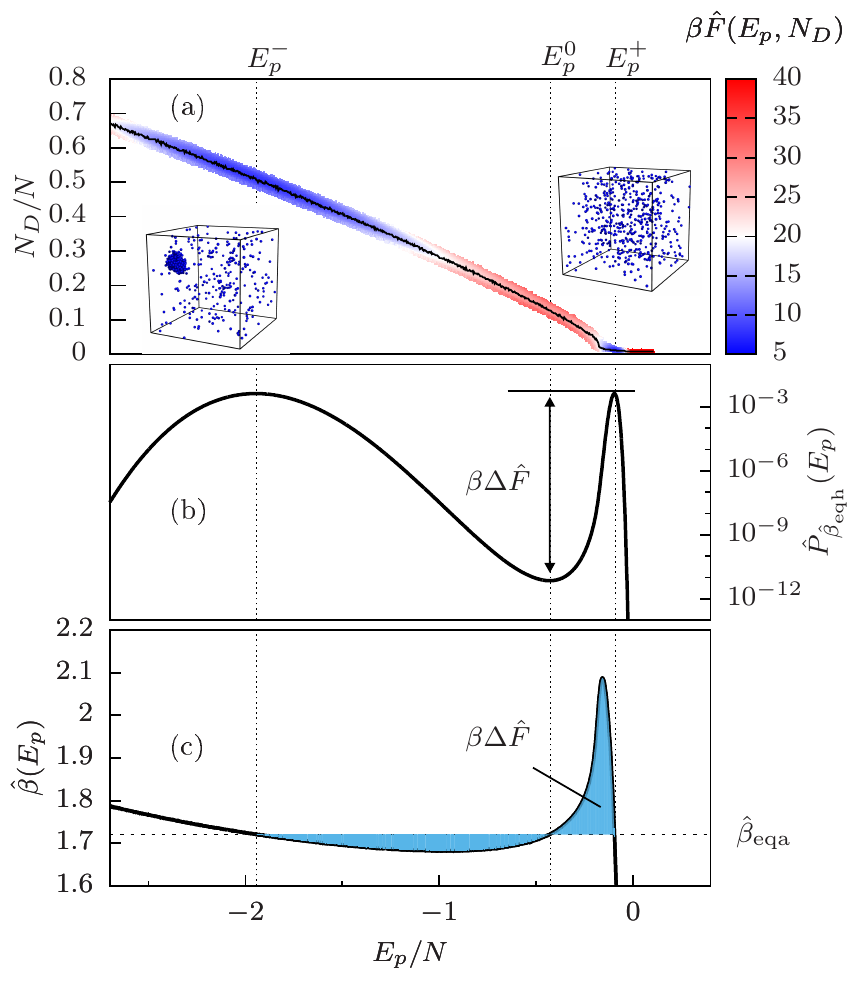}
  \vspace{-2em}
  \begin{flushleft}
    $\boldsymbol{E}$:
  \end{flushleft}
  \vspace{-1.8em}
  \includegraphics{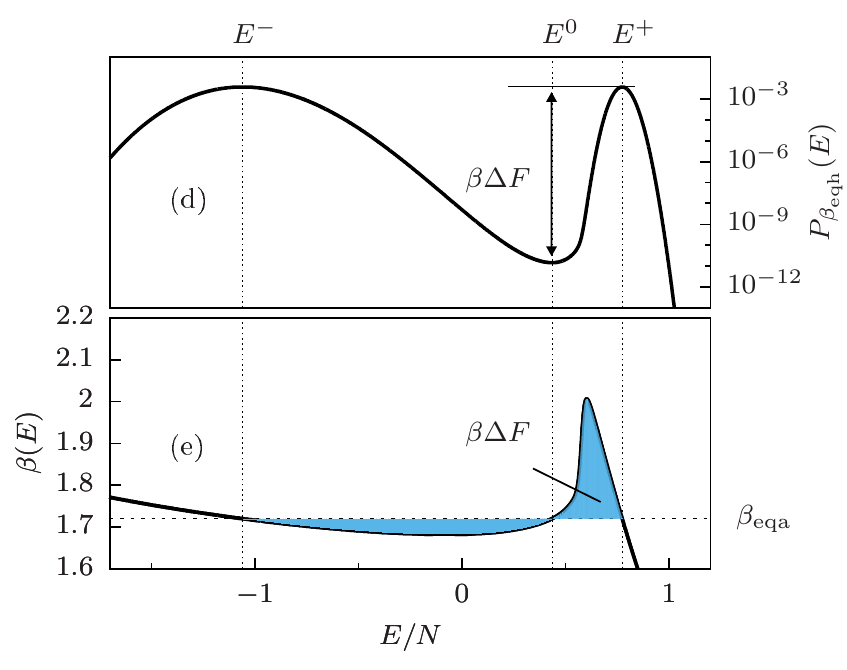}
  \vspace{-2em}
  \caption{%
    {\bf Free-energy barrier of droplet formation.}
    (a) Illustration of the free-energy landscape $\beta\hat{F}(E_p,N_D)$ (color
    map) as a function of potential energy $E_p$ and droplet size $N_D$ for
    \mbox{$N=512$} Lennard-Jones particles. The minimal free-energy path (black
    solid line) connects a droplet ($E_p\approx E_p^-$) and a gaseous
    ($E_p\approx E_p^+$) phase, visualized by the snapshots at $E_p^{\pm}$.
    (b) The projection onto the reaction-coordinate $E_p$ yields the canonical
    potential-energy probability distribution $\hat{P}_\beta(E_p)$, where the
    free-energy barrier $\beta\hat\Delta F$ is encoded in the ratio between
    maximum and minimum at $\betaconfeqh$. 
    (c) Equivalently, $\beta\hat\Delta F$ is the (equal) area size enclosed
    between the microcanonical inverse temperature $\hat\beta(E_p)$ and the
    accordingly defined transition temperature $\betaconfeqa$, where
    \mbox{$\betaconfeqa=\betaconfeqh=1.72099(3)$}. 
    The analogous quantities are reevaluated as a function of total energy $E$
    in (d) and (e) with \mbox{$\betatexteqa=\betatexteqh=1.71999(3)$}.
    \vspace{-1em}
  }
  \label{figIllustration}
\end{figure}
We begin with the paradigm of nucleation and dissolution, the equilibrium
droplet formation in a supersaturated particle gas~\cite{binder1980,
furukawa1982, neuhaus2003, biskup2002, biskup2003, binder2003, nussbaumer2006,
nussbaumer2008, zierenberg2015}.
%A paradigm of nucleation and dissolution is the equilibrium droplet formation in
%a supersaturated gas~\cite{binder1980,neuhaus2003, biskup2002, biskup2003,
%binder2003, nussbaumer2006, nussbaumer2008, zierenberg2015}.  
%theoretical expectations
The first-order condensation-evaporation transition separates a homogeneous gas
phase from an inhomogeneous phase, where a single macroscopic droplet of size
$N_D$ is in equilibrium with the remaining vapor~\cite{binder1980,
furukawa1982, biskup2002, biskup2003, neuhaus2003, binder2003}. In fact, the
probability for intermediate-sized droplets was shown to
vanish~\cite{biskup2002, biskup2003}.  In the vicinity of the transition, the
energy-dominated inhomogeneous condensed phase coexists with the
entropy-dominated homogeneous gas phase. A transition between both phases may
only occur by energy variation upon nucleation or dissolution. In reality, this
of course refers to the total energy $E$. For systems where the momentum phase
space may be integrated out explicitly (see Methods), the problem simplifies in
terms of computability. We hence begin with an illustration in the
potential-energy formulation (denoted by a hat, e.g., $\hat{F}$) as a direct
result of computer simulations before we go over to comparing both energy
approaches.

Figure~\ref{figIllustration}~(a) shows the free-energy landscape
$\beta\hat{F}(E_p, N_D)$ of droplet condensation-evaporation of Lennard-Jones
particles (see Methods) at the finite-size transition temperature. We define
\mbox{$\beta\hat{F}(E_p, N_D)=-\ln\left[\Omega(E_p,N_D)e^{-\beta E_p}\right]$},
where $\Omega(E_p,N_D)$ is a generalization of the (conformational) density of
states to the two-dimensional $E_p$--$N_D$ reaction-coordinate space.
% MEP
For $E_p$ fixed, $\beta\hat{F}(E_p, N_D)$ resembles a parabola with a single
local minimum. The resulting transition path is shown as a black line and its
relative maximum along this path (around $E_p^0$) is an estimate of the
transition free-energy barrier. 
% relation to suppression
Instead, however, one may consider the projection along the droplet
size~\cite{auer2001} or equivalently along the energy, connected to the
corresponding probability distributions
$\beta\hat{F}(N_D)=-\ln\hat{P}_\beta(N_D)$ and
$\beta\hat{F}(E_p)=-\ln\hat{P}_\beta(E_p)$.

We follow the latter approach and derive the free-energy barrier from the
suppression of transition states in the canonical energy probability
distributions (see Methods) for both reaction coordinates $E$ and $E_p$. 
The probability distributions are shown in Figs.~\ref{figIllustration}~(b) and
(d) and are clearly asymmetric, with a narrow peak for the gas phase and a
broad peak for the droplet phase. 
% barrier
Methodologically, both ensembles are analogous so that we limit in the following
the notation to the case of total energy $E$. At the equal-height inverse
temperature $\betatexteqh$, $P_{\betatexteqh}(E)$ has two peaks at $E^{\pm}$ of
equal height and in between a minimum at $E^{0}$. The resulting free-energy
barrier is 
%
%\begin{equation}
\mbox{$\beta\Delta F = \ln\left(P_{\betatexteqh}(E^{\pm})/P_{\betatexteqh}(E^{0})\right)$}.
%\end{equation}
%
Equivalently, one may perform the analysis entirely in the microcanonical
frame~\cite{gross2001} and consider the enclosed area by the microcanonical
inverse temperature $\beta(E)$ and the canonical inverse temperature (see Methods)
\begin{equation}
  \beta\Delta F = \int_{E^{0}}^{E^{\pm}}dE~[\beta(E) - \beta],
  \label{eqDeltaFBeta}
\end{equation}
shown in Figs.~\ref{figIllustration}~(c) and (e). Demanding areas of equal size
yields the equal-area inverse temperature $\betatexteqa$, which is in fact
identical to $\betatexteqh$~\cite{janke1998}.

We notice that the energy probability distributions $P_\beta(E)$ and
$\hat{P}_\beta(E_p)$ are related by a convolution involving the
Maxwell-Boltzmann distribution $P_{\rm MB}(x)$ as
\mbox{$P_\beta(E)=(\hat{P}_\beta\ast P_{\rm MB})(E)$} (see Methods). This
in turn corresponds to a physical smoothing, which diminishes the ratio between
maxima and minimum.  As a consequence we expect a lower barrier in the
total-energy formulation due to the kinetic contribution.

%%%%%%%%%%%%%%%%%%%%%%%%%%%%%%%%%%%%%%%%%%%%%%%%%%%%%%%%%%%%%%%%%%%%%%%%%%%%%%%%
%%%%%%%%%%%%%%%%%%%%%%%%%%%%%%%%%%%%%%%%%%%%%%%%%%%%%%%%%%%%%%%%%%%%%%%%%%%%%%%%
%%%%%%%%%%%%%%%%%%%%%%%%%%%%%%%%%%%%%%%%%%%%%%%%%%%%%%%%%%%%%%%%%%%%%%%%%%%%%%%%
%%%%%%%%% Results
%
%%%%%%%%%%%%%%%%%%%%%%%%%%%%%%%%%%%%%%%%%%%%%%%%%%%%%%%%%%%%%%%%%%%%%%%%%%%%%%%
%%%%%%%% expected scaling of the free-energy barrier
\subsection{Finite-size scaling of free-energy barrier}
\begin{figure}
  \vspace{-3em}
  \includegraphics[width=\columnwidth]{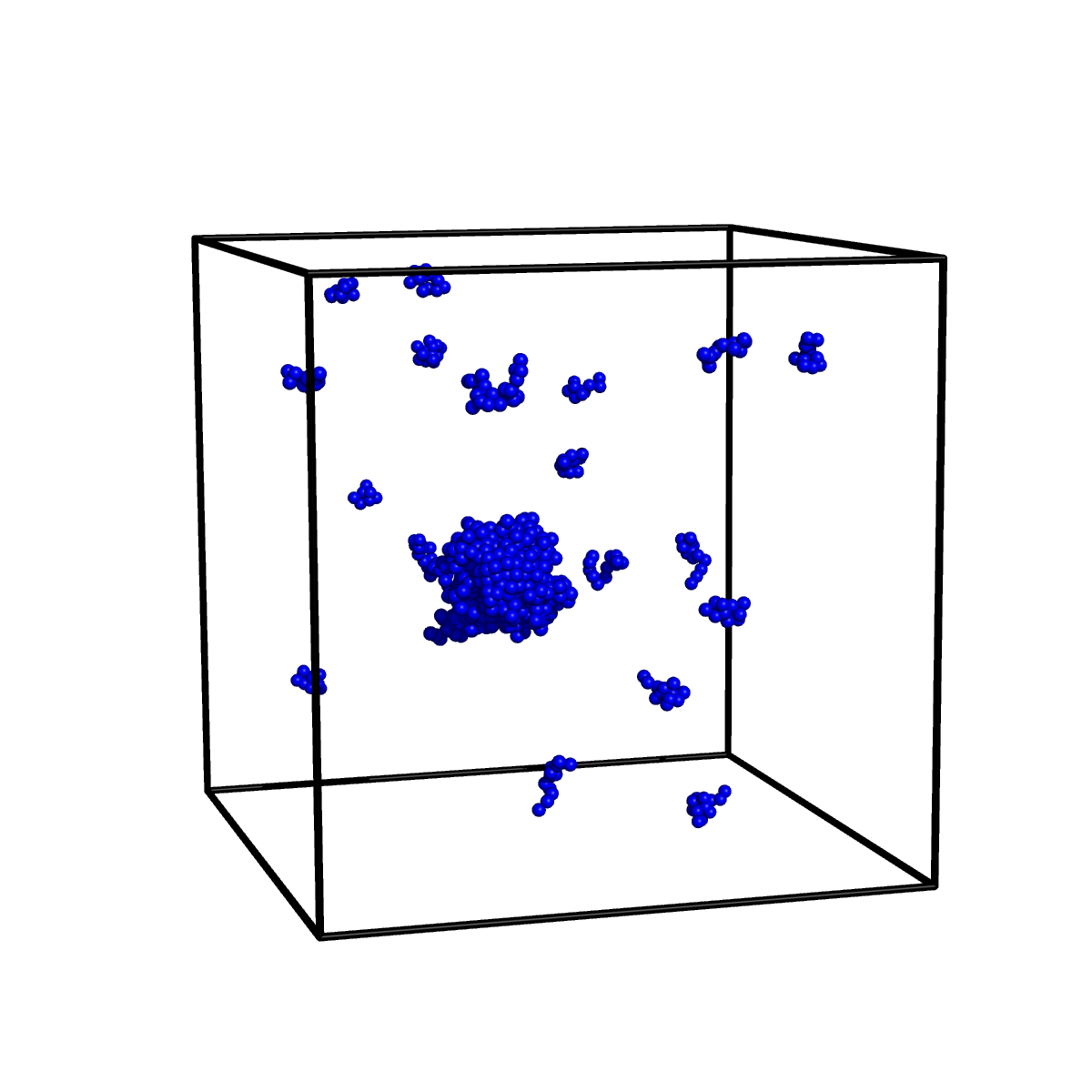}
  \vspace{-3em}
  \caption{%
    {\bf Polymer aggregate in a dilute solution.}
    Illustration of a cluster or aggregate of polymers in a dilute solution
    ($N=64$ bead-spring polymers with $13$ monomers each; monomer density
    $\rho=10^{-2}$). The snapshot stems from the droplet phase ($E_p\approx
    E_p^-$).
  }
  \label{figImagePoly}
\end{figure}

In the following, we discuss the free-energy barrier of droplet formation in a
particle gas and a dilute polymer solution as a function of system size. We
here extend the notion of droplet formation to the formation of clusters or
aggregates in polymeric systems. In particular, we consider linear bead-spring
polymers (see Methods), each consisting of $13$ monomers. The resulting polymer
cluster or aggregate is coexisting with a polymer ``gas'', see
Fig.~\ref{figImagePoly}, a first sign for the analogy to particle droplet
formation.  

The free-energy barrier is commonly assumed to be proportional to the occurring
interface. Here the surface of the droplet $\partial V_D$ separates the liquid
droplet from the surrounding gas and consequently we expect $\beta\Delta F =
\sigma \partial V_D$, with the interface tension $\sigma$.
% surface 
For any non-fractal shape, the surface area is related to the droplet volume $V_D$
as $\partial V_D \propto V_D^{2/3}$. Since nucleation shows no sign of critical
behavior, this is a physically valid assumption. 
% fixed temperature
However, at the condensation-evaporation transition $V_D$ itself does not scale
trivially with system size $V$. At a fixed temperature, general arguments
exploiting the equivalence to the Ising model imply that droplet formation is
triggered by insertion of particles until a single macroscopic droplet of size
$V_D \propto V^{3/4}$ coexists with the surrounding vapor~\cite{biskup2002,
biskup2003, neuhaus2003, binder2003}. 
% fixed density
This result may be translated to a
fixed-density setup using Taylor expansions, where directly at the finite-size
transition temperature the analogue scaling $V_D \propto N^{3/4}$ was verified
for both lattice and off-lattice particle models~\cite{zierenberg2015}.
 
Putting everything together and introducing an effective interfacial free energy
$\tauEff$ then yields to leading order
%
%\begin{equation}
  $\beta\Delta F  \propto \tauEff N^{1/2}$.
%  \label{fssFreeEnergy}
%\end{equation}
%
It is common for the study of interface tensions to consider logarithmic
corrections~\cite{ryu2010, nussbaumer2010, prestipino2012, prestipino2013},
dating back to early field-theoretic results~\cite{langer1967}. The physical
origin are translational invariance as well as capillary waves at the
interface. Altogether we use for our final scaling ansatz 
\begin{equation}
  \beta\Delta F = \tauEff N^{1/2} - \alpha\ln N + c,
  \label{fssFreeEnergy}
\end{equation}
where $\alpha$ and $c$ are constants. This is the leading-order exponent in
Eq.~\eqref{eqRate}. Neglecting the prefactor $\kappa$ for now, we obtain from
Eq.~\eqref{fssFreeEnergy} to leading order the rate of equilibrium droplet
formation as $R\propto N^{\alpha}e^{-\tauEff N^{1/2}}$.  Thus, for increasing
system size the probability that a single macroscopic droplet forms decreases
exponentially.

\begin{figure}
  \includegraphics{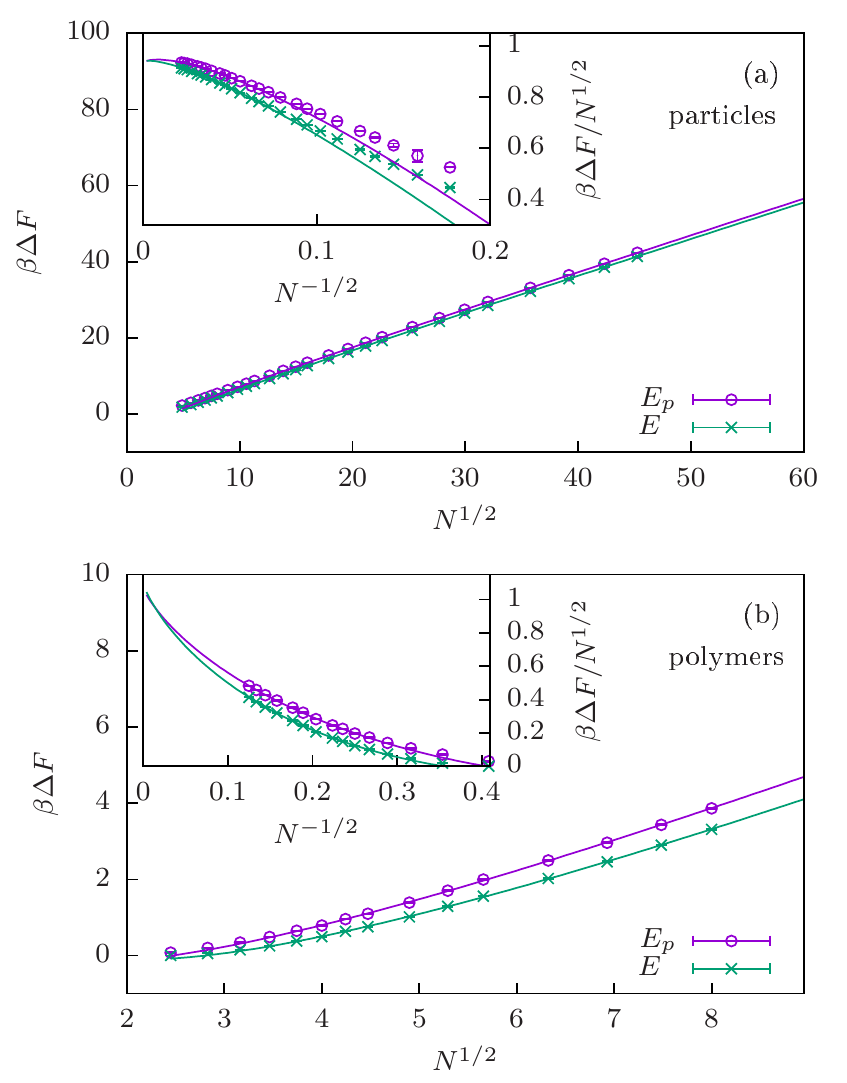}
  \caption{%
    {\bf Finite-size scaling of free-energy barrier.}
    Free-energy barriers $\beta\Delta\hat{F}$ (reaction coordinate $E_p$) and
    $\beta\Delta F$ (reaction coordinate $E$) of droplet formation in a particle
    gas (a) and dilute polymer solution (b) as a function of the number of
    constituents $N$. The leading $N^{1/2}$ scaling is clearly demonstrated. The
    insets show the finite-size scaling of the interfacial free energy according
    to Eq.~\eqref{fssFreeEnergy}. Error bars are obtained from Jackknife error
    analysis.
  }
  \label{figBarrier}
\end{figure}
%%%%%%%% particles
Figure~\ref{figBarrier}~(a) shows the free-energy barrier for droplet formation
in a particle gas as a function of system
size for both reaction coordinates $E$ and $E_p$, obtained via
Eq.~\eqref{eqDeltaFBeta}. Both estimates yield barriers up to about $42\kB T$,
showing at close sight an almost constant shift. Only the total energy $E$
includes the kinetic contribution which is here reflected in a smaller barrier,
whereas the interfacial free energies are expected to be identical,
$\tautext_{\rm eff}=\tauconf_{\rm eff}$.
% fit
In fact, fits to Eq.~\eqref{fssFreeEnergy} yield $\tautext_{\rm eff}=0.939(4)$
and $\tauconf_{\rm eff}=0.935(4)$, each for $N\geq N_{\rm min}=320$ with
goodness-of-fit parameter $Q\approx 0.3$, for reaction coordinate $E$ and $E_p$,
respectively. This remains consistent within error bars under variation of
$N_{\rm min}\in [224,1280]$. The accordance of data and fit is demonstrated in
the inset. 
In order to test the significance of the logarithmic corrections, we considered
in addition a restricted fit to $\beta\Delta F=\tauEff N^{1/2}+c$. We obtain
$\tautext_{\rm eff}=0.973(2)$ and $\tauconf_{\rm eff}=0.977(2)$ for $N_{\rm
min}=768$ with $Q\approx 0.2$ and $Q\approx 0.1$, respectively. However, the
estimate of $\tauEff$ gradually decreases with increasing $N_{\rm min}$.  Thus,
the most probable scenario remain small logarithmic corrections competing with
a constant shift.
% remark
The leading scaling behaviour was also observed in Ref.~\cite{statt2015} directly
as a linear function of the droplet surface. The advantage of the present
approach is that it avoids difficulties and uncertainties coming from the
``correct'' identification of the cluster surface.

%%%%%%%% polymers
For the cluster formation in a dilute polymer solution shown in
Fig.~\ref{figBarrier}~(b), the situation remains  qualitatively similar. Fits to
Eq.~\eqref{fssFreeEnergy} yield $\tautext_{\rm eff}=1.06(3)$ and $\tauconf_{\rm
eff}=1.03(3)$, each for $N_{\rm min} = 16$ with $Q\approx 0.2$, for reaction
coordinate $E$ and $E_p$, respectively. Note that this is on the same scale as
for droplet formation of particles in Fig.~\eqref{fssFreeEnergy}~(a).  Compared
to the particle case the system sizes are, however, much smaller, making
quantitative predictions for the polymer case less reliable. Still, the overall
behavior supports the hypothesis that cluster formation in a dilute polymer
solution shows a strong analogy to droplet formation in a particle gas.

% kinetic contribution
We observe that considering the kinetic contribution results in a shifted
barrier. For the considered examples and relevant system sizes the shift is
about $\beta\Delta\hat{F}-\beta\Delta F\approx0.5\dots1$, i.e., of the order
$\mathcal{O}(1)$. 
This additive contribution, while much smaller than the leading behaviour, leads
to a multiplicative relation between the nucleation rates $R\propto
e^{-\beta\Delta F}\approx e^{-(\beta\Delta \hat{F}-1)}\approx 3e^{-\beta\Delta
\hat{F}}\propto 3\hat{R}$.  Neglecting the momentum phase space thus
underestimates the rates. In common situations, however, the deviations between
experiment and simulations are of the order of several magnitudes, such that the
effect of the kinetic contribution may be considered subleading.

%%%%%%%%%%%%%%%%%%%%%%%%%%%%%%%%%%%%%%%%%%%%%%%%%%%%%%%%%%%%%%%%%%%%%%%%%%%%%%%
%%%%%%%% temperature
\subsection{Finite-size scaling of transition temperature}
\begin{figure}
  \includegraphics{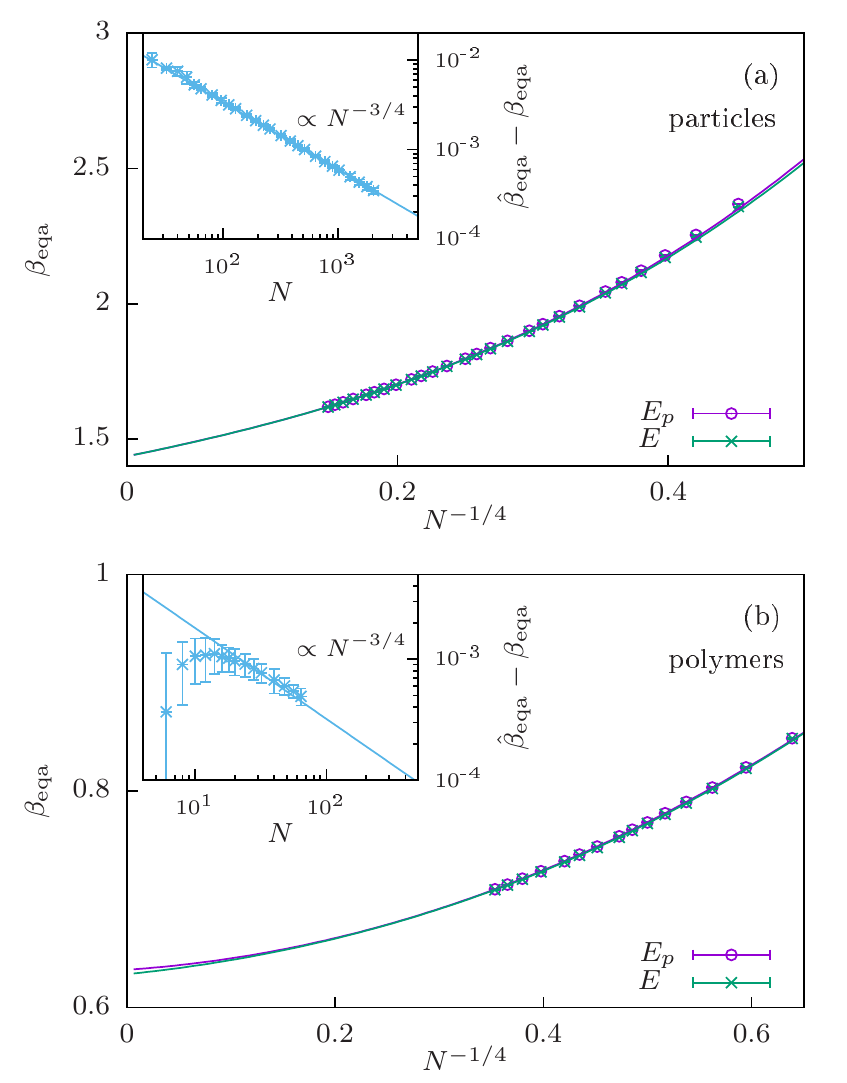}
  \caption{%
    {\bf Finite-size scaling of transition temperature.}
    Inverse transition temperature $\betaconfeqa$ (reaction coordinate $E_p$)
    and $\betatexteqa$ (reaction coordinate $E$) of droplet formation in a
    particle gas (a) and dilute polymer solution (b) as a function of the number
    of constituents $N$. The finite-size scaling ansatz Eq.~\eqref{eqAnsatzBeta}
    describes the data perfectly. The inset shows a vanishing finite-size
    difference between both ensembles $\betaconfeqa(N)-\betatexteqa(N)\propto
    N^{-3/4}$. Error bars are obtained from Jackknife error
    analysis.
    %Finite-size scaling of the transition temperature for droplet formation in a
    %particle gas and dilute polymer solution considering reaction
    %coordinates $E_p$ and $E$. The inset shows a vanishing finite-size
    %difference between both temperatures.
    \vspace{-0.5em}
  }
  \label{figFSSTemp}
\end{figure}
The evaluation of the free-energy barrier via equal areas provides us with a
definition of the finite-size transition temperature. At fixed density, we
showed for the condensation-evaporation transition~\cite{zierenberg2015} that
the transition temperature, obtained from specific-heat peaks, scales as inverse
power of the critical droplet radius $R_D\propto V_D^{1/3}\propto N^{1/4}$. The
same is expected for all other transition temperature definitions.
Figure~\ref{figFSSTemp} shows the equal-area definition together with a fit
including higher-order corrections of the form
\begin{equation}
  \betatexteqa(N) = \beta_0 + a N^{-1/4} + b N^{-1/2} + c N^{-3/4},
  \label{eqAnsatzBeta}
\end{equation}
for cluster formation in both particle gas and polymer solution.
%%%%%%%% particles
In the case of particle condensation, least-square fits for $N_{\rm
min}=192$ yield $\betatext_0=1.436(2)$ and $\betaconf_0=1.436(2)$ each with
$Q\approx0.5$, for reaction coordinate $E$ and $E_p$, respectively.
The excellent fit results show that the empirical, yet physically motivated,
higher-order corrections describe the finite-size scaling very well.
% exp motivation
In addition, the strong finite-size deviations open a possibility to study
finite-size scaling in experiments on the nanoscale (for a conversion see
Methods). The finite-size transition temperature of the largest system
($N=2048$) is $\betatexteqa\sim1.61948(2)$, which still deviates from the
thermodynamic limit by $\mathcal{O}\left(10\%\right)$.

It is worth noting that typical canonical finite-size transition temperatures,
e.g., the peak location of the specific heat, do not depend on the kinetic
contribution to the partition function. In the canonical expectation values, the
kinetic prefactor simply cancels. Here, however, we observe a finite-size
difference in the transition temperature depending on whether we consider the
kinetic contribution or not. Of course, the thermodynamic limit coincides. This
is illustrated in the inset of Fig.~\ref{figFSSTemp}~(a) with a finite-size scaling
of the transition temperature difference $\betaconfeqa(N)-\betatexteqa(N)$. It
shows a prominent power-law scaling of the form $\sim N^{-3/4}$, which
interestingly is the same scaling as the inverse transition droplet volume. The
finite-size difference arises from the convolution of an asymmetric energy
probability distribution with the Maxwell-Boltzmann distribution, compare
Fig.~\ref{figIllustration} and Eq.~\eqref{eqDistribution} in Methods, which
manifests in the geometric differences of the microcanonical inverse temperature
and the enclosed areas in Eq.~\eqref{eqDeltaFBeta}. It appears that correlations
between the ensemble definitions account for compatible leading-order scaling
corrections in Eq.~\eqref{eqAnsatzBeta}, which further supports this ansatz and
explains the observed difference.

%%%%%%%% polymers
For polymer aggregation in Fig.~\ref{figFSSTemp}~(b), fits of
Eq.~\eqref{eqAnsatzBeta} yield $\betatext_0=0.64(2)$ and $\betaconf_0=0.64(2)$
for $N_{\rm min}=14$ (guided by the inset) each with $Q\approx0.8$, for reaction
coordinate $E$ and $E_p$, respectively. Qualitatively, the fit ansatz describes
the data already well when including the smallest system sizes. Also the
finite-size ensemble deviation in the inset shows a clear $N^{-3/4}$ trend for
$N\geq14$. Again, this is an indication for the analogy between cluster
formation in polymer solutions and droplet formation in a particle gas.

%% comparison to saddle-piont
%This is not in contrast to the pointwise finite-size difference of the two
%microcanonical inverse temperatures $\sim N^{-1}$ from a saddle-point
%approximation (see Methods), because $\betatexteqa$ and $\betaconfeqa$ are
%obtained as a solution to Eq.~\eqref{eqDeltaFBeta}, which results in a
%nontrivial relationship.
%%
%Instead, it appears that correlations between the ensemble definitions account
%for compatible leading-order scaling corrections in Eq.~\eqref{eqAnsatzBeta},
%which further supports this ansatz and explains the observed difference.

%%%%%%%%%%%%%%%%%%%%%%%%%%%%%%%%%%%%%%%%%%%%%%%%%%%%%%%%%%%%%%%%%%%%%%%%%%%%%%%%
%%%%%%%%%%%%%%%%%%%%%%%%%%%%%%%%%%%%%%%%%%%%%%%%%%%%%%%%%%%%%%%%%%%%%%%%%%%%%%%%
%%%%%%%%%%%%%%%%%%%%%%%%%%%%%%%%%%%%%%%%%%%%%%%%%%%%%%%%%%%%%%%%%%%%%%%%%%%%%%%%
%%%%%%%%% Conclusions
\section{Discussion}
%Restricting the discussion to the conformational phase space, considering as
%reaction coordinate $E_p$ instead of $E$, is a valid approach for the estimation
%of canonical expectation values. However, we have shown that the absolute
%free-energy barrier shows a marginal finite-size difference which leads to a
%multiplicative factor in the nucleation rate. This may become relevant once
%theoretical predictions and experimental measurements becomes precise enough. 
%%
%On the other hand, a finite-size scaling analysis of the interfacial free energy
%and the related transition temperature revealed that intensive properties
%coincide in the thermodynamic limit. To some extent, this resolves the ensemble
%ambiguity and justifies the consideration of the conformational ensemble. Still,
%physical interpretations about energy transfer and kinetics are only accessible
%in the total-energy formulation. The additional information comes from the exact
%knowledge of the kinetic contribution and is more than a reformulation of the
%problem at hand. For example, microcanonical negative temperatures are not
%possible in the $NVE$ ensemble. 
%%
%As a numerical advantage, considering the total energy as reaction coordinate
%shows less fluctuations since the convolution of the microcanonical partition
%function and canonical probability distribution is a smoothing procedure of
%physical origin.

%droplet condensation
We have presented a shape-free approach to the estimation of canonical
free-energy barriers in equilibrium droplet formation. The finite-size scaling
is dominated by the  predicted $N^{1/2}$ behaviour but we identified additional
logarithmic corrections from precise numerical estimates.
% kinetic contributions
Somewhat surprisingly, the absolute free-energy barrier is sensitive to the
consideration of the kinetic contribution. It is well-known that the restriction
to the conformational phase space does not influence finite-size transition
points determined from canonical expectation values. These are evaluated on the
level of the canonical partition function. The free-energy barrier and the
associated equal-height or equal-area transition temperature, however, are
determined from the energy probability distribution. Here, the two formulations
are related by a convolution with the Maxwell-Boltzmann distribution which
explains the finite-size differences. At the same time, the probability
distributions are the integrands of the respective partition functions.  In the
end, this boils down to the trivial fact that equality of integrals does not
imply equality of the integrands.
% implications
This may become relevant once theoretical predictions and experimental
measurements become precise enough. Still, a restriction to the conformational
phase space retains intensive parameters in the thermodynamic limit.
As a numerical advantage, considering the total energy as reaction coordinate
leads to less fluctuations in the microcanonical partition function and
canonical probability distribution, since the underlying convolution is a
smoothing procedure of physical origin.

We provided evidence that the derived finite-size scaling of canonical droplet
formation is applicable to cluster formation in dilute polymer solutions as
well, despite the a priori non-trivial shape of the polymer cluster. This is a
clear indication of an analogy between particle condensation and polymer
aggregation. In particular, we showed that polymer clusters are in equilibrium
with non-attached (free) polymers: an inhomogeneous or mixed phase of aggregate
and solute polymers.  
This is intuitively clear when the polymers are interpreted as extended
particles.
%an observation that has been previously exploited modeling peptide aggregation
%by a lattice gas with asymmetric interactions~\cite{irbaeck2015}. 
The leading-order corrections then follow from the interplay between energy
minimization by forming a local cluster and entropy maximization by retaining
freely movable constituents. Of course, additional corrections should follow
from the explicit geometry and internal behavior of the constituents. 

% generalization
It is expected that the energy remains a suitable reaction coordinate for
general nucleation-like mechanisms. In this case, generalized-ensemble methods
may unfold their full power. Moreover, our approach at fixed density provides
the possibility for experiments to perform heating-cooling studies in order to
probe transition rates. 
The presented results for polymer aggregation suggest that this approach may be
generalized to studies of protein cluster formation~\cite{stradner2004}. In a
wider scope, it may also find potential application for temperature-driven
self-assembly~\cite{jonkheijm2006, huisman2008}, crystallization in
phase-change materials~\cite{kalb2005, wuttig2009, lee2011} and glassy
solids~\cite{lee2009}, dislocation nucleation~\cite{ryu2011}, or the study of
surface nanobubbles and nanodroplets~\cite{lohse2015}.
% heterogenity
Of course, experimental observations commonly include the formation of multiple
clusters. Reasons for this include heterogeneities or impurities acting as
nucleation seeds. We suppose that this leads to a local quasi-equilibrium on the
respective length scales. Here, a proper combination of the canonical droplet
formation with the effect of nucleation seeds~\cite{auer2004} seems to be a
fruitful approach to a systematic understanding.
With further developments, simulations may provide reliable estimates for
finite-size systems and meet experiments on the nanometer scale.

%%%%%%%%%%%%%%%%%%%%%%%%%%%%%%%%%%%%%%%%%%%%%%%%%%%%%%%%%%%%%%%%%%%%%%%%%%%%%%%%
%%%%%%%%%%%%%%%%%%%%%%%%%%%%%%%%%%%%%%%%%%%%%%%%%%%%%%%%%%%%%%%%%%%%%%%%%%%%%%%%
%%%%%%%%%%%%%%%%%%%%%%%%%%%%%%%%%%%%%%%%%%%%%%%%%%%%%%%%%%%%%%%%%%%%%%%%%%%%%%%%
%%%%%%% Model/Method setup
\section{Methods}
\subsection{Microcanonical ensembles}
Recently, there has been some ambiguity with the definition of a
``microcanonical ensemble'' in computer simulations~\cite{schierz2015}. This is
a crucial aspect relevant for physical interpretations that appears to be
unwittingly softened in the last decade. 
%%%%% textbook
The \emph{microcanonical ensemble} ($NVE$) describes an isolated system in which
the number of constituents $N$, the volume $V$, and the total energy $E$ are
conserved. Here, the transfer of potential energy $E_p$ into kinetic energy
$E_k$ and vice versa is a valid and relevant mechanism, where $E=E_k+E_p$.
% entropy 
The microcanonical (Boltzmann) entropy is defined as 
$\Stext(E)=\kB\ln\Gamma(E)$,
with the partition function 
\mbox{$\Gamma(E) = \int\int\mathcal{D}x~\mathcal{D}p~\delta(E-[E_p(x)+E_k(p)])$},
where $\mathcal{D}x$ denotes the integration over state space and $\mathcal{D}p$ over momentum space.

%%%%% conformational
The other common definition is the \emph{conformational microcanonical ensemble}
($NVE_p$), describing instead a system with fixed potential energy $E_p$. The
conformational microcanonical entropy is
$\Sconf(E_p)=\kB\ln\Omega(E_p)$, 
where
$\Omega(E_p)=\int \mathcal{D}x~\delta(E_p-E_p(x))$
is the density of states or the conformational microcanonical partition
function. The consequences are drastic: A (physical) interpretation of energy
transfer from potential to kinetic energy is no longer valid. 
%importance
This is natural for spin systems, where a kinetic contribution is not defined in
the first place (but may be exploited for numerical
purposes~\cite{martinmayor2007}).  On the contrary, it is particularly relevant
for situations in soft condensed matter, e.g., for particles and polymers, where
interpretations of energy transfer become natural. However, there are good
reasons for this choice: 
$\Omega(E_p)$ is a fundamental property of statistical mechanics. It encodes the
full information about the conformational space and allows for identification of
(structural) phase transitions~\cite{gross2001, janke1998, kastner2009,
schnabel2011}. Furthermore, it may be exploited for reweighting techniques and
flat-histogram Monte Carlo methods~\cite{janke2012, janke2016,
zierenberg2016polymers}.

%%%%% relation
The relation between $\Gamma(E)$ and $\Omega(E_p)$ is given by a convolution
with the kinetic-energy contribution (see, e.g., Ref.~\cite{labastie1990}): If momenta and
positions are independent one may separate the kinetic-energy contribution
$E_k=\sum_{i}p_i^2/2m$, explicitly perform the momentum integration, and
obtain for $N$ particles in three dimensions~\cite{calvo2000}
\begin{equation}
  \Gamma(E)=\frac{(2\pi m)^{\frac{3N}{2}}}{\Gamma(\frac{3N}{2})}\int_{-\infty}^{\infty} dE_p
  \Omega(E_p) (E-E_p)^{\frac{3N-2}{2}}\Theta(E-E_p).
  \label{eqGamma}
\end{equation}
We define \mbox{$\Omega(E_p)=0\quad\forall\quad E_p<E_{p, {\rm min}}$} in order
to extend the integral over the full energy range $(-\infty,\infty)$. In this
way the total-energy-surface entropy $S(E)$ appears as a (weighted)
potential-energy-volume entropy. Notice that, since all $\Omega(E_p)>0$, the
classical $NVE$ entropy increases with $E$ and the microcanonical inverse
temperature
\mbox{$\kB\betatext(E)=\partial \Stext(E)/\partial E$}
cannot become negative, opposed to its conformational counterpart
\mbox{$\kB\betaconf(E_p)=\partial \Sconf(E_p)/\partial E_p$}.
This may be an interesting aspect for a recent debate on the correct definition
of entropy when connected with the phenomenological thermodynamic entropy, e.g.,
in Refs.~\cite{dunkel2014, hilbert2014, swendsen2015} and references therein.

%define how we really measure beta(E) and beta(E_p)!!!
Numerically, we determine the microcanonical inverse temperatures as follows.
In the conformational microcanonical ensemble, we have direct access to an
estimate of $\ln\Omega(E_p)$ (see below) such that $\beta(E_p)$ is obtained by a
numerical 5-point derivative. In the full microcanonical ensemble, we may
estimate the inverse temperature in terms of microcanonical expectation values
for $N$ independent particles
\begin{equation}
   \langle O\rangle_E
   =\frac{\int dE_p~O(E_p)~\Omega(E_p) (E-E_p)^{\frac{3N-2}{2}}\Theta(E-E_p)}{\int dE_p~\Omega(E_p)
   (E-E_p)^{\frac{3N-2}{2}}\Theta(E-E_p)},
\end{equation}
where the explicit prefactor in Eq.~\eqref{eqGamma} cancels. Then, we may
express $\beta(E) = \partial \ln\Gamma(E)/\partial E = \frac{3N-2}{2}\langle
\frac{1}{E-E_p}\rangle_{E}$.

%todo: carry over to long-version
%While this is, in principle, an advantage for a well-defined measurement of
%free-energy barriers, we found it in many classical nucleation-related cases
%only relevant for very small system sizes or model-dependent artificial energy
%regimes.

%todo: carry over to long-version
% include saddle-point approximation
%\newcommand{\Epmin}{E_p^{(0)}}
%\newcommand{\epmin}{e_p^{(0)}}
%
%The energies $E$ and $E_p$ may be related by identifying the maximum of the
%integrand $\exp[f_E(E_p)]$ in Eq.~\eqref{eqGamma}, which is the $NVE$
%potential-energy distribution~\cite{schierz2015}. This is unique if the
%distribution has a single peak and corresponds to the maximum $\Epmin$ of
%\mbox{$f_E(E_p)=\Sconf(E_p)/\kB+[(3N-2)/2]\ln(E-E_p)$}, which has to satisfy
%%
%\begin{equation}
%  (E-\Epmin)/N = e-\epmin = \left(\frac{3}{2}-\frac{1}{N}\right)
%  \betaconf^{-1}(\Epmin).
%  \label{eqMin}
%\end{equation}
%%
%The energy-per-particle difference may be identified with an average kinetic
%energy \mbox{$e-\epmin = e_k$}, which converges to the prediction of the
%equipartition theorem with increasing system size. Moreover, the leading term of
%a saddle-point approximation around $\Epmin$ allows to relate the microcanonical
%inverse temperatures 
%\mbox{$\betatext(E)= \betaconf(\Epmin) + \mathcal{O}\left(N^{-1}\right)$},
%which restores the ensemble equivalence in the thermodynamic limit wherever
%$\Epmin$ is unique.

\subsection{Canonical ensembles}
The \emph{canonical ensemble} is defined in terms of the partition function 
$Z_{\beta} = \int\int \mathcal{D}x~\mathcal{D}p~e^{-\beta E}$,
where each phase-space point is weighted with the Boltzmann factor according to
the total energy. Again, the kinetic part may be explicitly integrated for
generic systems. Each degree of freedom contributes with a Gaussian integral, and
one obtains for $N$ particles, 
\begin{equation}
  Z_{\beta} = (2\pi m/\beta)^{3N/2} \hat{Z}_\beta,
  \label{eqRelationZ}
\end{equation}
where 
$\hat{Z}_{\beta} = \int \mathcal{D}x~e^{-\beta E_p}$
is the partition function of the \emph{conformational canonical ensemble}. 
Both partition functions may be expressed as integrals in terms of the
respective energies, namely \mbox{$Z_\beta=\int dE~\Gamma(E)e^{-\beta E}$} and
\mbox{$\hat{Z}_\beta=\int dE_p~\Omega(E_p)e^{-\beta E_p}$}. The corresponding canonical
energy probability distributions are defined as
\mbox{$P_{\beta}(E)=\Gamma(E)e^{-\beta E}/Z_\beta$} and
\mbox{$\hat{P}_{\beta}(E_p)=\Omega(E_p)e^{-\beta E_p}/\hat{Z}_\beta$}.

We may now relate the two energy probability distributions by starting with the
definition of $P_\beta(E)$ and inserting Eq.~\eqref{eqGamma} and
Eq.~\eqref{eqRelationZ}:
\begin{widetext}
\begin{equation}
P_{\beta}(E)\\=\int_{-\infty}^{\infty} dE_p
\hat{P}_{\beta}(E_p)\frac{\beta^{\frac{3N}{2}}}{\Gamma(\frac{3N}{2})}(E-E_p)^{\frac{3N-2}{2}}e^{-\beta(E-E_p)}\Theta(E-E_p).
\label{eqDistribution}
\end{equation}
\end{widetext}
We identify the latter part of the integrand as the $N$-particle
Maxwell-Boltzmann distribution $P_{\rm MB}(x)$ and may write
Eq.~\eqref{eqDistribution} as a convolution
\mbox{$P_{\beta}(E)=(\hat{P}_{\beta}\ast P_{\rm MB})(E)$}. 

% Lennard-Jones
\subsection{Lennard-Jones particles}
We consider a system of Lennard-Jones particles in a dimensionless periodic box
of length $L$ with fixed density \mbox{$\rho=N/L^3=10^{-2}$}. Mutual avoidance
and short-range attraction are modeled by the 12-6 Lennard-Jones potential 
\mbox{$V_{\rm LJ}(r) = 4\epsilon \left[ (\sigma/r)^{12} - (\sigma/r)^6 \right]$}
with $\epsilon=1$ and $\sigma=2^{-1/6}$ such that $r_{\rm min}=1$. 
For computational efficiency, the potential is cutoff at $r_c=2.5\sigma$ and
shifted by $-V_{\rm LJ}(r_c)$. System sizes range up to $N=2048$, which is
competitive with state-of-the-art Molecular Dynamics simulations like
well-tempered metadynamics~\cite{salvalaglio2015}. For the chosen density
$\rho=10^{-2}$ a system with $2048$ particles requires a box of linear dimension
$L'\approx59~r'_{\rm min} = 59\times~2^{1/6}~\sigma'$. For argon,
$\sigma'\approx3.4\mathring{A}$ such that $L'\approx22.5~{\rm nm}$ is on the
nanoscale. Of course, for a comparison to an experimental setup one should
include both the explicit geometric constraints and the full Lennard-Jones
potential. 

% Bead-Spring
\subsection{Bead-spring polymers}
The considered dilute polymer solution is modeled by $N$ linear bead-spring
polymers, consisting of $13$ monomers each, again in a dimensionless periodic
box with monomer density \mbox{$\rho=13N/L^3=10^{-2}$}. Bonds are modeled
between neighboring monomers by the FENE potential 
\mbox{$V_{\rm FENE}(r)=-(KR^2/2)\ln[1-(r-r_0)^2/R^2]$} with $K=40$, $R=0.3$, and
$r_0=0.7$.
Non-bonded monomers interact with the same Lennard-Jones potential as above but
with $\sigma=r_0~2^{-1/6}$ such that $r_{\rm min}=r_0$~\cite{zierenberg2014jcp,
zierenberg2015epl, zierenberg2016polymers}.
% convert to some scale (maybe polyethylene) evtl. 0.38nm
% Proc Math Phys Eng Sci. 2015 Aug 8; 471(2180): 20150171. 
% simga=0.38nm=r_0 2^{-1/6} =0.7*a*2^{-1/6}
% a=0.38nm/0.7*2^{1/6}=0.6093nm
% L=43.66a = 26.6nm
%
% degrees of freedom
The total number of monomers is $13N$, which yields $3\times13N$ total momentum
degrees of freedom in Eq.~\eqref{eqGamma} and successive relations. The bounded
bond length \mbox{$[r_0-R,r_0+R]$} from the FENE potential formally introduces
constraints on these degrees of freedom. However, for practical applications in
ordinary temperature ranges this effect is negligible and reweighting to the
full microcanonical and canonical ensemble is feasible~\cite{schierz2015}.

% muca
\subsection{Multicanonical Monte Carlo simulations}
Parallel multicanonical Monte Carlo simulations~\cite{zierenberg2013,
zierenberg2014, berg1991, berg1992, janke1992, janke1998b} allow us to
efficiently sample the suppressed states, by iteratively adapting an auxiliary
weight function $W(E_p)$ to yield a flat histogram $H(E_p)$. The final weight
function is related to the density of states up to a multiplicative factor:
\mbox{$\Omega(E_p)\propto H(E_p)/W(E_p)$}. This gives direct access to
microcanonical estimates and canonical expectation values at any temperature.
Using Eq.~\eqref{eqGamma} this even provides an estimate of $\Gamma(E)$. 
%details 
Monte Carlo updates for the particle case include short-range and long-range
particle displacements. For updates of the polymer configurations, we employed
local single-monomer shifts, bond-rotation and double-bridging moves, as well as
long-range polymer displacements
We measure the conformational (potential) energy $E_p$ and the number of
particles in the largest cluster $N_D$ as in Ref.~\cite{zierenberg2015}. Error
bars are obtained by the Jackknife method~\cite{efron1982, young2015}.

\subsection{Data availability}
The data that support the findings of this study are available from the
corresponding author upon request.

\section{References}
%\section*{References}

%%%%%%%%%%%%%%%%%%%%%%%%%%%%%%%%%%%%%%%%%%%%%%%%%%%%%%%%%%%%%%%%%%%%%%%%%%%%%%%
%%%%%%%%% Acknowledgments
\section{acknowledgments}
The project was funded by the European Union, the Free State of Saxony and the
Deutsche Forschungsgemeinschaft (DFG) under Grant No.\ JA~483/31-1 and
Sonderforschungsbereich/Transregio SFB/TRR 102 (Project B04). The authors
gratefully acknowledge the computing time provided by the John von Neumann
Institute for Computing (NIC) on the supercomputer JURECA at J\"ulich
Supercomputing Centre (JSC) under Grant No.\ HLZ24. 
Part of this work has been financially supported by the Leipzig Graduate School
of Natural Sciences ``BuildMoNa'' and by the Deutsch-Franz\"osische Hochschule
(DFH-UFA) through the Doctoral College ``${\mathbb L}^4$'' under Grant No.\
CDFA-02-07.

\vspace{1em}
\section{Contributions}
All authors contributed equally to the theory and J.Z. performed the
simulations. J.Z. and W.J. analyzed the data and wrote the paper. All authors
discussed the results and commented on the manuscript.

\section{Competing financial interests}
The authors declare no competing financial interests.

\end{document}